# Polymer-inspired mechanical metamaterials

Zhenyang Gao[a,b,c,d], Pengyuan Ren[a,b], Yifeng Dong[d], Gengchen Zheng[a,b], Minh-Son Pham[e], Xiao Shang[c], Shaojia Wang[c,f], Shuo Yang[f], Zijue Tang[a,b], Yongbing Li[g,h], Hua Sun[a,b], Yi Wu[a,b,i*], Hongjian Jiang[a,b], Lan Zhang[j], Tobin Filleter[f], Lingyu Kong[k], Kun Zhou[d], Haowei Wang[a,b,i,l], Yang Lu[m], Yu Zou[c*], Hongze Wang[a,b,i,n*]

[a]State Key Labortory of Metal Matrix Composites, Shanghai Jiao Tong University, Shanghai, 200240, China
[b]School of Materials Science and Engineering, Shanghai Jiao Tong University, Shanghai, 200240, China
[c]Department of Materials Science and Engineering, University of Toronto, Toronto, ON M5S 3E4, Canada
[d]School of Mechanical and Aerospace Engineering, Nanyang Technological University, Singapore, 639798, Singapore
[e]Department of Materials, Imperial College London, London, SW7 2AZ, UK
[f]Department of Mechanical & Industrial Engineering, University of Toronto, Toronto, ON M5S 3G8, Canada
[g]Shanghai Key Laboratory of Digital Manufacture for Thin-walled Structures, Shanghai Jiao Tong University, Shanghai, 200240, China
[h]State Key Laboratory of Mechanical System and Vibration, Shanghai Jiao Tong University, Shanghai, 200240, China
[i]Institute of Alumics Materials, Shanghai Jiao Tong University (Anhui), Huaibei, 235000, China
[j]Research Center for Intelligent Robotics, Zhejiang Lab, Hangzhou, 311100, China
[k]Institute of Marine Equipment, Shanghai Jiao Tong University, Shanghai, 200240, China
[l]Anhui Province Industrial Generic Technology Research Center for Alumics Materials, Huaibei Normal University, Huaibei, Anhui 235000, China
[m]Department of Mechanical Engineering, University of Hong Kong, Hongkong, 999077, China
[n]Shanghai Key Laboratory of Material Laser Processing and Modification (Shanghai Jiao Tong University, School of Materials Science and Engineering), Shanghai 200240, China

Corresponding authors: eagle51@sjtu.edu.cn (Yi Wu)
mse.zou@utoronto.ca (Yu Zou)
hz.wang@sjtu.edu.cn (Hongze Wang)

**Abstract**

Metamaterials benefit from unique architected patterns to achieve lightweight with exceptional mechanical properties inaccessible to conventional materials. Typical mechanical metamaterials are inspired by crystal-like lattice structures, whose closely packed frameworks often exhibit a rigid mechanical nature. Here, we present polymer-inspired metamaterials (PIMs) by programming deformation and strengthening mechanisms that mimic the mechanical roles of key constituent elements in polymer networks. By combining metamaterial programmability with polymer-inspired structures, we design crosslinking, proto-crystalline order, and entanglement in PIMs to enable macroscale strengthening mechanisms inspired by crosslink, molecular-density, and pre-stretch strengthening in polymers, expanding the metamaterial structure-property design space. This macroscale polymer-inspired programmability also suggests that PIMs could serve as a design platform incorporating the programmability strategies to achieve desired deformation and strengthening responses, holding a potential for applications in soft robotic joints and compliant connectors.

## 1. Introduction

Mechanical metamaterials[1, 2, 3], created by novel, intricate internal microstructures, exhibit mechanical properties beyond natural materials. Strategies for internal microstructure are mainly inspired by crystal-inspired structures and bonding like body-centered-cubic (BCC)[4], face-centered-cubic (FCC)[5], or octet-truss[6] to obtain superior mechanical characteristics such as exceptional stiffness[7, 8, 9], strength[10, 11], damage tolerance[12, 13, 14], and energy absorption[15, 16, 17]. However, the internal architecture, relying on either the crystal-like periodic arrangements, or employing the straight connection (e.g. struts or plates) similar to atomic bonding, exhibits a rigidity that limits plastic yielding, stretchability and fracture strains and is heavily dependent on the inherent properties of parent materials to expand their attainable property space. To broaden the material property space of mechanical metamaterials, existing studies have extensively explored new topological designs[18, 19, 20, 21, 22] and introduced different strengthening mechanisms[13, 23, 24, 25] for these crystal-inspired metamaterials (CIMs). Some research has also focused on biomimetic structural modifications[16, 26, 27, 28] or working on their parent materials[29, 30, 31, 32] to overcome this limit. Nevertheless, the fundamental deformation mechanics of structures with crystal-inspired bonding remain unchanged, which limits the expansion of their material properties and, consequently, their application range. In contrast, polymers, with their twisted, entangled, and soft molecular chains reinforced by crosslinks, exhibit a high level of plasticity and molecular mechanics different from that in crystals[33, 34]. This raises a fundamental question: can mechanical metamaterials be engineered to mimic polymer-enabled behaviors, thereby changing the deformation mechanics of existing metamaterial designs?

Replicating polymer microstructures[35, 36, 37] in metamaterials, characterized by highly twisted and crosslinked molecular chains, remains challenging. This complexity is beyond regular atomic arrangements in crystal materials[38, 39, 40]. Notable studies[41,42] have transformed CIM-derived structures into twisted architectures to achieve enhanced flexibility and functional performance[43,44], such as fracture resistance[45], large extendability[46], extreme specific strength[47], and highly recoverable deformation[48]. These advantages inspire the development of more stretchable architected beams that are mechanically similar to flexible polymer-chain elements. Building on these features, it is of interest to move toward truly polymer-network-based structures, in which chain-based architectures are interconnected by crosslinks. In addition, it is also meaningful to design polymer-inspired chains with non-periodic and more flexible architectures to enable polymer-inspired mechanical behaviour at the architectural scale. For polymer-inspired networks, a successful translation of polymer strengthening mechanisms, including molecular density (MD) strengthening, crosslink (CL) strengthening, and pre-stretch (PS) strengthening, into metamaterial design will enable the development of new type of metamaterials having polymer-like mechanical behaviour for functional opportunities.

Here, we developed the polymeric structure generation (PSG) algorithm to create the polymer-inspired metamaterials (PIMs) with programmable polymer-inspired chains and crosslinks that mimic the mechanical strengthening found in the polymer molecular structures (Fig. S28, Supplementary Movie 1). This design inspired from the typical mechanical effects of polymer networks, in which soft chains accommodate deformation through reorientation, contour change, and progressive load transfer, while crosslinks restrict chain motion and stabilize the network. PIM displays polymer-inspired elastic, plastic, and fracture behaviors in macroscale, filling the gap in materials Ashby charts that are currently not achievable by CIMs. Most interestingly, a programmability can be achieved for PIMs thanks to architectural features that enable CL, PS, and MD strengthening. To demonstrate potential for applications, we created programmable humanoid tissues with PIMs, showcasing engineered robotic toes and fingers that mimic human mechanics. In addition, we combined inspirations from crystals and polymers to create composite CIM+PIM with significant anisotropic mechanics enabling soft biometric exoskeletons and multimodal grabs with highly stretchable and energy-absorbing features.

## 2. Results and discussion

### *2.1. Polymer-inspired meta-structures and programmable mechanics in metamaterials*

In contrast to crystal-inspired meta-structures (Fig. 1a), metamaterials inspired by polymers needs to have architectural features that mimicking complex molecular structures of twisted, entangled soft chains stabilized by crosslinks (Fig. 1b). To mimic such complex structures, we developed the PSG algorithm (Supplementary Note 1) to generate curved, twisted chain features that are cross-linked by straight ligaments. While the curved chains are to emulate the compliant and deformable mechanics of polymer molecular chains, the straight crosslinks connect chains to constrain the chain motion. These PIM architectural features can be described by parameters including the number of polymer-inspired chains packed per unit volume (molecular-like density, $\rho$), proportion of length of chains aligned with load direction (PS level, $p$), and the number of crosslinks per unit volume (crosslink density, $n_c$ ), Fig. 1c. Programming these parameters enables translation of chain entanglement, directional alignment, and crosslink-constrained deformation to create PIMs that have strengthening mechanisms inspired by those in polymers: (1) PS strengthening, which uses pre-aligned and partially ordered polymer-inspired chains along the loading direction to increase specific strength and stiffness; (2) MD strengthening, which exploits polymer-inspired chain entanglement within a confined volume to either restrict chain mobility and improve failure resistance or permit extremely large elastic deformation for greater flexibility; and (3) CL strengthening, which increases specific stiffness by restricting the mechanical deformation among neighboring polymer-inspired chains. Therefore, we can study to obtain in-depth understandings, then subsequently program these molecular strategies to address a question: can polymer strengthening mechanisms be translated to benefit mechanical metamaterials, and can metamaterials, in turn, provide a programmable platform to interrogate these mechanisms at a higher level of abstraction?

While the deformation mechanics of polymers at the molecular scale are well established, an equivalent framework for polymer-like deformation in architected metamaterials at the macroscale is still largely missing. This gap prevents us from combining the structural programmability of metamaterials with polymer mechanics to create artificial materials with programmable polymer-inspired functionalities. To program PIMs, the deformation and strengthening behaviour of a "polymer chain" at the macroscale must first be understood. To do so, we modelled an architectural chain inspired by polymer molecular chains, as shown in Fig. 2a, then compared the macroscale mechanical properties with the intrinsic polymer strengthening, subsequently use macroscale energy partition modelling to interpret the architectural deformation mechanisms and program the properties (see Supplementary Note 6 for details). Digital image correlation (DIC) and mechanical testing (Fig. 2b and Supplementary Movie 2) reveal macroscale responses of polymer-inspired strengthening features in PIMs. In PIMs, entanglements, PS segments and crosslinked polymer-inspired chains locally constrain chain motion, resulted in elevated local strain captured by DIC maps in Fig. 2b.

Increasing molecular density promotes the chain entanglement ($\nu_{ent}$), which restricts local mobility and concentrates strain within entangled zones by redirecting the energy partition from reorientation-dominated configurations toward chain-bending modes with lower reorientation energy partition ($\Phi_{reorient}$) and higher bending energy partition ($\Phi_{bend}$). PS chains in PIMs align earlier under load, increasing the load carried per unit global strain but at the same time reducing the available non-affine accommodation and thus promoting a stiffer, less extensible response once debonding mode becomes dominant. Crosslinked regions ($\nu_{cl}, f_{cl}$) constrain the extension of polymer-inspired chain, suppressing reorientation and bending modes with reduced $\Phi_{reorient}$ and $\Phi_{bend}$ and generating localized high-strain, debonding-dominated fields where the debonding energy partition ($\Phi_{debond}$) becomes the primary component of mechanical work (refer to Supplementary Note 6 for the model relations between the energy partitions and these strengthening events). Finite element analysis (Fig. 2c and Fig. S21–S25) is used to quantify how entanglements per chain ($\nu_{ent}$), crosslinks per chain ($\nu_{cl}$), proto-crystalline ordering ($\Omega$), and crosslink functionality ($f_{cl}$) reshuffle the energy partitions (see Supplementary Note 6 for detailed and quantitative derivations).

To program the polymer-inspired strengthening in PIMs, the relations between the macroscopic energy-partitions and mechanical properties of MD, CL, and PS strengthening features designed by PSG algorithms were studied. We obtain architectures whose measured modulus $E$, strength $\sigma_c$, and fracture energies and strains follow the programmed responses with quantitative agreement between prediction and experiment (Fig. 2d, see Supplementary Note 6 and Fig. S31-32 for held-out validation of model). We also experimentally compare the measured responses of polymer-inspired strengthening with intrinsic polymer strengthening, where the modulus $E$ and strength $\sigma_c$ are proportional to the effective network chain density $\nu_e$ and $\sqrt{\nu_e}$, respectively, and engineering relations of polymer pre-stretch gives power-law increases of $\lambda^{n_E}$ and $\lambda^{n_\sigma}$ with draw ratio $\lambda$ (see Supplementary Note 7 for detailed derivations and references). These results establish a programmable mechanics framework that translates polymer strengthening mechanisms into architected metamaterials and links design parameters directly to macroscopic PIM energy partitions and mechanical responses.

*2.2. Programming polymer strengthening to surpass nature*

Guided by the polymer-inspired strengthening mechanisms, our results demonstrate that PIMs can be programmed to go beyond the current structural property limits (Fig. 3). Here, we demonstrate the capability of PIM through programming to improve targeted mechanical properties, including specific modulus (Fig. 3a), specific strength (Fig. 3b), energy absorption efficiency (Fig. 3c), and fracture behaviors (Fig. 3d), in comparison with CIMs (with and without twisted structures), plate-based lattices, shell-based TPMS gyroids, and directional 2D lattices (see Fig. S29 and Supplementary Note 2 for detailed design schematics and information). The PIM designs include CL, PS, MD strengthening, inverse PS and inverse MD (inverse PS+MD), PS with high MD (PS+MD), spatially programmed PS with high MD (programmed PS+MD), and PS, high-density crosslinked chains (PS+CL+MD).

PIMs programmed with PS macro-chains and maximized artificially engineered molecular densities were designed to favour debonding-dominated crosslinked behaviour, achieving an increase of 8 times in specific modulus (Fig. 3a). Meanwhile, PIMs combining polymer-inspired crystalline PS strengthening (PS (crystalline)) with MD strengthening exhibit a more than 4 times increase in specific strength. (Fig. 3b, also see Supplementary Note 8 for crystalline PIM design). Conversely, a PIM with inverse MD and PS design is programmed to enhance chain-bending and reorientation contributions, increasing plastic plateau and resulting in an increase by 65% in compressive energy-absorption efficiency despite of a reduction in weight (Fig. 3c). By prescribing PS distributions together with high MD to target gradient ($\Phi_{debond}$, $\Phi_{bend}$, $\Phi_{reorient}$) triplets, PIMs are programmed to exhibit a more ductile and progressive fracture process, absorbing 4x higher fracture energy and dissipating 5x more energy during fracture propagation relative to twisted CIMs. (Fig. 3d). Our polymer-inspired strengthening approach also allows program the mechanical behaviour precisely along given orientations, including crack control (Supplementary Note 5), leading to a new type of metamaterials that incorporates the mechanisms in polymers to significantly enhance and program the mechanical response, including efficient energy absorption, ultra-high modulus and strength at low weight, and improved fracture resistance. To compare the fundamental building elements of PIMs with those of strut-based metamaterials at the individual strut scale, we experimentally measured the maximum fracture strain of PIM chain elements, classical straight struts, and different helical struts under identical beam diameter and design-space conditions. Experiments show that a 1 cm PIM chain built with durable v2 resin can be drawn up to 54.8 cm, resulting in a 5,480% maximum fracture strain (Supplementary Movie 4) with a specific energy absorption of 4.8 J/g, where the fracture strain can be further extended to 6,730% if fabricated by flexible 80A resin. For a broad range of parent materials with intrinsic fracture strains $\varepsilon_f = 12\%$ (clear v4), $\varepsilon_f = 42\%$ (durable v2), and $\varepsilon_f = 63\%$ (flexible 80A), PIM chain increased the fracture strains by 48, 25, and 18 times, respectively, compared to the helical strut elements. These results provide an elementary-scale basis for the exceptional extendibility and fracture dissipation of PIMs.

Mechanistically, in crystal-inspired lattices, straight members experience stretching/bending along well-constrained load paths, where axial-stretching along struts leads to early yielding and abrupt failure. In PIMs, the architected chains first spend a large portion of the deformation in reorientation- and chain-bending–dominated regimes, where deformation is accommodated by twisting, uncoiling, and contour reshaping rather than immediate backbone stretch. Designs such as PS+MD and PS (crystalline)+MD deliberately suppress these responses and maximise the debonding partition $\Phi_{debond}$ over high-density chains parallelly programmed, which amplifies stiffness and strength. In contrast, the inverse PS+MD architecture promotes $\Phi_{reorient}$ and $\Phi_{bend}$ over a wide strain window, creating an extended plastic-like plateau and higher energy absorption. Artificial PS+MD distributions imprint spatially varying ($\Phi_{debond}$, $\Phi_{bend}$, $\Phi_{reorient}$) triplets to program a smoothly evolving resistance rather than abrupt transtion between deformation regiums, giving stable, high-energy fracture dissipation. Finally, the "single-chain" zero-entanglement PIM pushes the entire structure through reorientation-dominated deformation before significant debonding occurs, drastrically increasing the fracture strain by orders of magnitude.

*2.3. Programming polymer-inspired metamaterials for humanoid tissues*

Biological muscles and tendons play a pivotal role in linking bones and transmitting forces due to their flexible, but tough mechanical properties. These properties are essential for the development of soft robotic joints that emulate human mechanics. Here, we have engineered humanoid tissues using PS toughened PIMs (Supplementary Note 3) that feature different polymer-inspired chain strain factors ( $f_c$ ) enabling programmable mechanics under external stimuli and diverse engineering load conditions:

$$f_c = k_{f_c}\rho^{b_E}p^{d_E}(n_c^{c_E} - 1) \quad (1)$$

where $b_E$, $d_E$, $c_E$ are experimentally calibrated exponents that quantify the sensitivity of the chain strain factor to molecular-like density, chain alignment, and crosslinking, respectively. $k_{f_c}$ is the physical constant related to the properties of parent materials. The DIC results (Fig. 4a, Fig. S13a-b, and Supplementary Movie 3) demonstrate that these humanoid tissues display a programmable strain response from stress stimulus, which escalated at higher levels of pre-stretching alignments. To tailor the mechanics of these humanoid tissues, we derived a framework for modulating the stress-responsive tissue modulus and $f_c$ under different PS levels (Fig. S13c). The elastic modulus of the tissue can be precisely programmed following a relationship outlined in equation (2), which is a derivation from the polymer-inspired chain strain from equations (1):

$$E = k_h f_c \quad (2)$$

where $k_h$ is a fitting constant related to the property of parent material. Enabled by workings of tendons and muscles, we illustrate two typical mechanics in human joints: (1) a typical walk cycle enabled by the toe (Fig. 4b), characterized by three stages with distinct reaction slopes ($k_1$, $k_2$, $k_3$), and (2) the load responses of actuated (Fig. 4c) and passive (Fig. S14) finger gestures with highly customized bending mechanics. Here we programmed the PIM tissues to develop robotic components that can replicate the human mechanics and being mechanically programmable. We developed PIM robotic foot joints with different gradient of polymer-inspired chain strain factors ($\nabla f_c$), where the higher $\nabla f_c$ resulted in less pre-stretching levels towards the edge of the joint and is expected to produce less stress reaction to the walk stimulus. This feature enables the programming of the humanoid toe mechanics. The robotic finger features a distributed polymer-inspired chain strain factor ($f_c$), tailored to emulate the bending reaction of a human finger. Mechanical experiment simulating the toe flexion in a complete walking cycle indicates a successful mimicry of the human foot mechanics (Fig. S34a-b, also see Fig S33c for a printed robotics feet). Observations suggest the smaller reaction slope magnitudes in samples with higher $\nabla f_c$ aligning with our design expectations, and informing design guidelines for programmable load reaction in response to walking stimuli:

$$k_i = K_i \nabla f_c + A_i \quad (3)$$

where $i$ = 1, 2, and 3 represents the indices of different reaction slopes, $K_i$ and $A_i$ are the experimentally derived parameters. Robotics humanoid fingers can be programmed as both actively actuated functional grabbing unit (Fig. 4c) or passive tissue joints being bended under external stresses. For active finger control, we developed a cable-driven robotic finger assembly, readily programmed through tailored distributions of different PIM joints. Experimental results confirmed that, by programming the $f_c$ distribution within the cable-driven finger, it could be actively actuated into various gestures, enabling highly customizable grabbing tasks. For passive finger designs, experiments simulating various human finger bending gestures (Fig. S14) demonstrate that the robotic finger successfully replicated the bending reactions of a human finger. This response can be precisely programmed by adjusting the polymer-inspired chain strain factor ($f_c$) for each flexion (Fig. S16). These results underscore the effectiveness of PIM and its programmable humanoid tissues, highlighting their potential for broad applications in biomimetic components that demand highly programmable and soft mechanics.

PIMs also demonstrated their capability to organically integrate diverse functional robotic and engineering structures in a programmable manner. For instance, composite materials incorporating metal particles within their polymer networks enhance mechanical properties while maintaining lightness, and can achieve anisotropically-tailored behaviors by strategically distributing these metal particles, a capability that was not reported before. Hence, we implanted crystal-inspired octet-truss lattices into the polymer networks (Supplementary Note 4), and adjusted crosslink densities, to create highly anisotropic CIM+PIM functional material sheets (Fig. 4d and Fig. S11), while it should be noted that our method is not confined to octet-truss structures or sheet formats. These composite CIM+PIMs enables applications in soft biometric exoskeletons (Fig. 4e) and cable-driven robotic grab fingers (Fig. 4f). Cable-driven active deformation of the grabber is illustrated in Fig. S35a, and the corresponding schematic is shown in Fig. S35b. The grabber is twisted by cable to conform to the object geometry, thereby generating multiple holding points for secure grabbing, while grab tension and friction act together to prevent object slippage or escape. The CIM+PIM, with its variable crosslink densities, enhances a programmable specific modulus along the sheet compression direction ($\vec{\boldsymbol{p}}$), while achieving an 85% reducible modulus at flexible stretch direction ($\vec{\boldsymbol{s}}$) compared to conventional metamaterials (Fig. S17a). It also exhibits a notable improvement in energy absorption capacity along $\vec{\boldsymbol{p}}$ over its CIM counterparts. Mechanical testing confirmed that the printed exoskeleton, featuring intricate carapaces interconnected with PIM and CIM under each carapace, responds with marked anisotropy to reshaping and shell-compression loads (Fig. S17b). This feature provides functional protection for soft robots, making them flexible and conformal for various functional modes as depicted in Fig. 1b. Additionally, the multimodal robotic finger, tested for soft bending and significant grab loads (Fig. S18), demonstrates the capability to deform effectively into complex shapes while exerting sufficient grabbing force to handle objects, as illustrated in Fig. 4f.

## 3. Conclusions

In this study, we introduce mechanical metamaterials inspired by polymer molecular structures. The PIM approach is enabled by programming polymer-like twisted chains crosslinked by straight beams to translate polymer-inspired strengthening mechanisms (MD strengthening, PS strengthening, CL strengthening) to meta-structures in PIMs. The developed PIMs demonstrate combinations of high specific strength, specific modulus, energy absorption efficiency, and ductile fracture behaviors, expanding the range of properties that are currently affordable in metamaterials using crystal-like stiff architectures. Our findings suggest that the mechanical response of PIMs can be rationalized through an architecture-level analogue of polymer strengthening mechanics. By synergistically combining crystal-inspired approach with the PIM one, we enable the highly anisotropic crystal-polymer-composite metamaterials that could be used in soft robotics, such as a biomimetic exoskeleton and a versatile grab finger. By programming the microscale mechanics of PIM chains, advances have been made towards creating programmable humanoid tissues that successfully simulate human-tissue mechanical properties. This research not only pioneers a new class of mechanical metamaterials inspired by polymers but also opens avenues for exploring polymer-like strengthening through the pre-programmed architectural design.

## 4. Methods

### *4.1. Materials and sample preparation*

In this paper, Formlabs® Form 3 stereolithography printer is applied to fabricate the samples. The parent materials in Fig. 2f are Formlabs® clear v4, durable, and flexible resins, where the durable resins are selected as the base materials for the rest of the figures due to its pliable but stiff mechanics simulating the properties of the molecular chains. It should be noted that the proposed method is not limited to these types of materials. The majority of PIMs are 3D printed by stereolithography (SLA) using Formlabs® durable resin to mimic the high stretchability of each polymer chains, while the clear v4 and flexible 80A resins are also used to systematically compare the mechanical improvement of PIMs for different parent materials. Experimental evidence confirms the first incorporation of the polymeric mechanics in metamaterials, characterized by molecular alignment, polymer-inspired chain scission, and extended fracture processes during the structural deformation. All samples are ultrasonically washed within absolute ethanol solutions for 10 min to remove adhered resin. The green clear v4 samples are tested without post-curing to avoid the excessive brittleness. The durable samples are cured in a 60 °C ultraviolet environment for 60 min, while the flexible samples are cured in a 60 °C ultraviolet environment for 10 min instead. All samples are secured in a dark environment and tested within 1 hours after the printing or post-processing. The mechanical properties of the parent materials are evaluated based on the ASTM D638 standard following the test configurations specified in Section 2 in Methods. The modulus, strength, fracture strain of different parent materials are summarized in Table S1.

Table S2 details the geometrical specifications and depicts various metamaterials, respectively. The diameters of the chains and bonding struts in PIMs, PIM-CIM composites, and PIM-based humanoid tissues have been standardized at 1 mm. Conversely, the diameters of the CIMs vary between 0.6 and 1.2 mm, facilitating a range of densities that overlap with those of the PIMs, enabling equitable comparisons in Ashby plots. The lower limit of the diameter ensures the print fidelity for these intricate geometries according to a voxel-based, additive manufacturing-oriented design verification approach introduced in our previous studies [41]. It is also important to note that the scope of the proposed method extends beyond these ranges of the structural design parameters for different fabrication technologies.

*4.2. Mechanical experiments*

Fig. S3 summarizes the test configurations used in this study, employing a ZwickRoell Z100 universal tensile test machine. Table S2 details the load rates, directions, and geometries for various mechanical tests satisfying quasi-static load conditions. ASTM D638 standard guidelines informed the mechanical evaluation of parent materials, with specimens conforming to ASTM D638 type IV (Fig. S3). PIM samples featuring pre-programmed strengthening mechanisms and CIM designs were tested in a 20 mm×10 mm×20 mm region along the x, y, and z axes, subject to vertical tensile or compressive forces (Fig. S4). Tensile testing for PIM+CIM composite layers focused on polymer chain loading, employing specimens with two PIM+CIM cells aligned with the load direction and one cell perpendicular, reducing necking phenomena and enhancing the generalizability of mechanical properties (Fig. S5(a)). Compressive tests for crystal loading mechanics used a single PIM+CIM cell perpendicular to the load direction(Fig. S5(b)). Biometric shells underwent tensile and compressive testing to assess reshaping and compression mechanics (Fig. S5(c-d)). A single grabbing finger was tested for both stretching (grab force) and bending (adapting force) (Fig. S5(e-f)). Cylindrical tissue test specimens with a molecular-like density of 4.25 $cm^{-3}$ under different PS levels were designed to evaluate tissue mechanics (Fig. S6(a)). For robotic toe and finger joints, specialized fixtures simulated bending loads akin to human tissue dynamics (Fig. S6(b-c)).

The experimental stress-strain curves are used to calculate the modulus, strength, energy absorption efficiency, fracture dissipation, and fracture energies of the specimens. The energy absorption efficiency is calculated by integrating the compressive stress-strain curve till the densification strain normalized by the peak stress (equation (6-7)). The fracture energies of evaluated by integrating the tensile stress-strain curves till the total fracture of the specimen, while the fracture dissipation integrates the curves from the fracture initiation strain to the complete fracture (equation (8-9)). The specific mechanical properties are derived by normalizing these mechanical performances by the density of the specimen within the test region to conduct fair comparisons.

$$\eta = \int_0^{\varepsilon_d} \sigma \varepsilon d\varepsilon / \sigma_{peak} \quad (4)$$

$$\sigma_{peak} = MAX\big(\sigma(\varepsilon)\big), \varepsilon \in [0, \varepsilon_P], \quad (5)$$

$$A_{fracture} = \int_0^{\delta_b} F \, d\delta, \quad (6)$$

$$A_{dissipation} = \int_{\delta_a}^{\delta_b} F \, d\delta, \quad (7)$$

where $\eta$ is the energy absorption efficiency, $\sigma$ is the compressive stress, $\sigma_{peak}$ is the peak stress, $\varepsilon$ is the compressive strain, $\varepsilon_P$ is the plateau initiation strain, $A_{fracture}$ is the fracture energy, $A_{dissipation}$ is the fracture dissipation energy, $F$ is the fracture load, $\delta_a$ and $\delta_b$ are the displacement at the fracture initiation and complete fracture, respectively.

*4.4. Digital image correlation analysis*

The DIC analysis is performed on a correlated Solutions VIC-3D DIC measurement system using its 2D measuring configurations (Fig. S7). The DIC is used to measure full-field surface strains during mechanical testing of PIM samples. These spatial strain maps were used to locate regions associated with different strengthening events (entangled zones, crosslinked segments, proto-crystalline chains) and quantify the relationship between experimental local chain strain and the programmed PIM design such as robotic muscles. The speckle patterns are sprayed evenly on the specimens for DIC calculations. A baser acA4112-30 um camera with 28 mm baser lens are used to capture the relative positions of the DIC speckle patterns, while VIC-2D 7 software is used to perform DIC analysis based on the experimental results. The analyzed images possess a resolution of 4096 pixels×3000 pixels. A 19 pixels×19 pixels subset dimension is used during the DIC derivation, while the step size is 4 pixels.

**Acknowledgment**
This work is sponsored by National Natural Science Foundation of China (523B2048, 52075327 and 52004160), Shanghai Sailing Program (20YF1419200), Natural Science Foundation of Shanghai (20ZR1427500), SJTU Global Strategic Partnership Fund (2023 SJTU-CORNELL), the innovation foundation of Commercial Aircraft Manufacturing Engineering center of China (No. 3-0410300-031), and the University Synergy Innovation Program of Anhui Province (GXXT-2022-086). Y.L. acknowledges the support from the Research Grants Council of the Hong Kong Special Administrative Region, China, under RFS2021-1S05; Hong Kong RGC general research fund #11200623; RGC Hong Kong under the CRF project C7074-23GF.

## Figures and tables

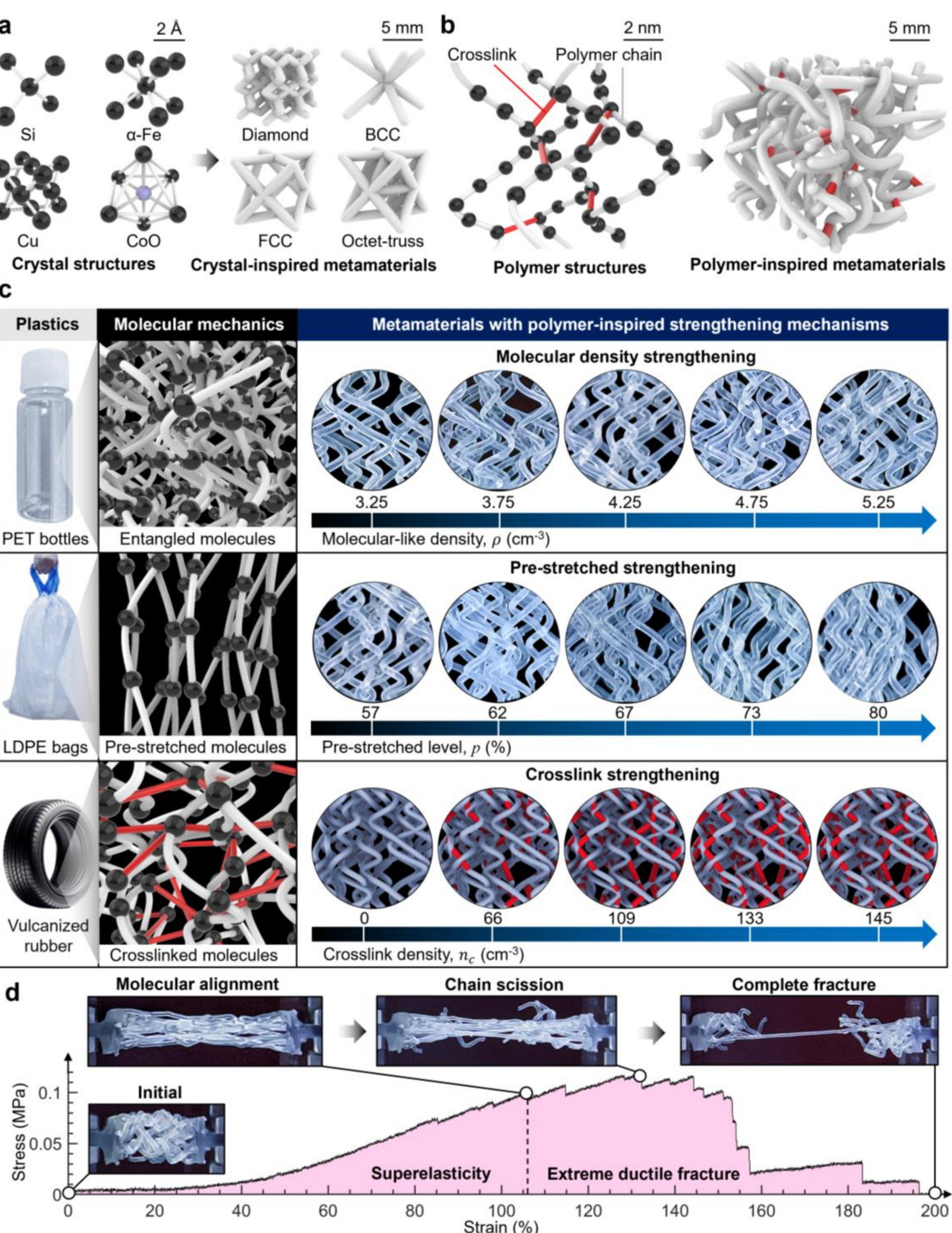

Fig. 1. The existing CIMs and proposed PIMs achieving different polymeric strengthening mechanisms. (a) Typical crystal structures and existing designs of CIMs. (b) The polymer structures and PIM. (c) The PIMs mimicking MD strengthening, PS strengthening, and CL strengthening inspired from the microstructural strengthening mechanisms of polymer. (d) Schematic PIM stress–strain response and representative deformation sequence.

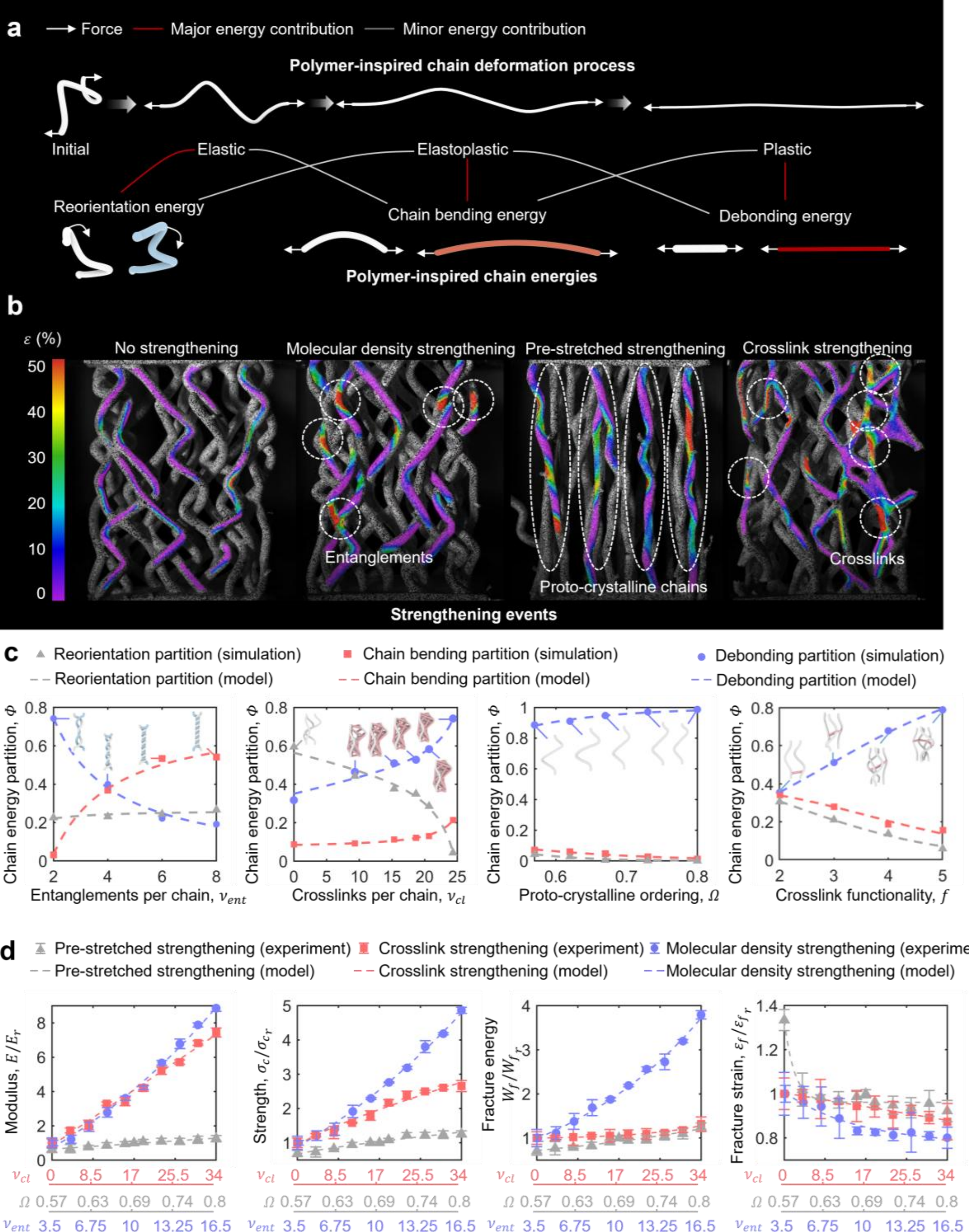

Fig. 2. Deformation mechanics and energy partitions of metamaterial mimicking polymer chains under different strengthening events, and theoretical modeling of chain energy partition and mechanical properties of polymer inspired metamaterials (PIMs). (a) Typical deformation process of a “polymer chain” in a PIM and the associated bonding energy partitions. (b) Strain fields from digital image correlation at 15% strain for a comparison specimen with $\rho$ = 4.25 cm$^{-3}$, a PIM with MD strengthening ($\rho$ = 5.25 cm$^{-3}$), a PIM with pre stretched strengthening ($p$ = 80%), and a PIM with CL strengthening ($n_c$ = 145 cm$^{-3}$).

Here $\rho$, $p$, and $n_c$ represent the molecular density, PS level, and crosslink density of the PIMs, respectively. (c) Chain energy partitions from different strengthening events obtained by finite element modeling and compared with predictions from the energy partition model. (d) Experimentally measured mechanical properties of PIMs and predictions from the macroscale polymer strengthening design model for different strengthening contributions. Here, $E_r$, $\sigma_{c_r}$, $W_{f_r}$, and $\varepsilon_{f_r}$ are normalization modulus, strength, fracture energy, and fracture strain for experimental measure at $\nu_{cl}$ = 0, $\Omega$ = 0.57, and $\nu_{ent}$ = 10. The error bars are obtained as standard deviations along three repetitive experimental measurements.

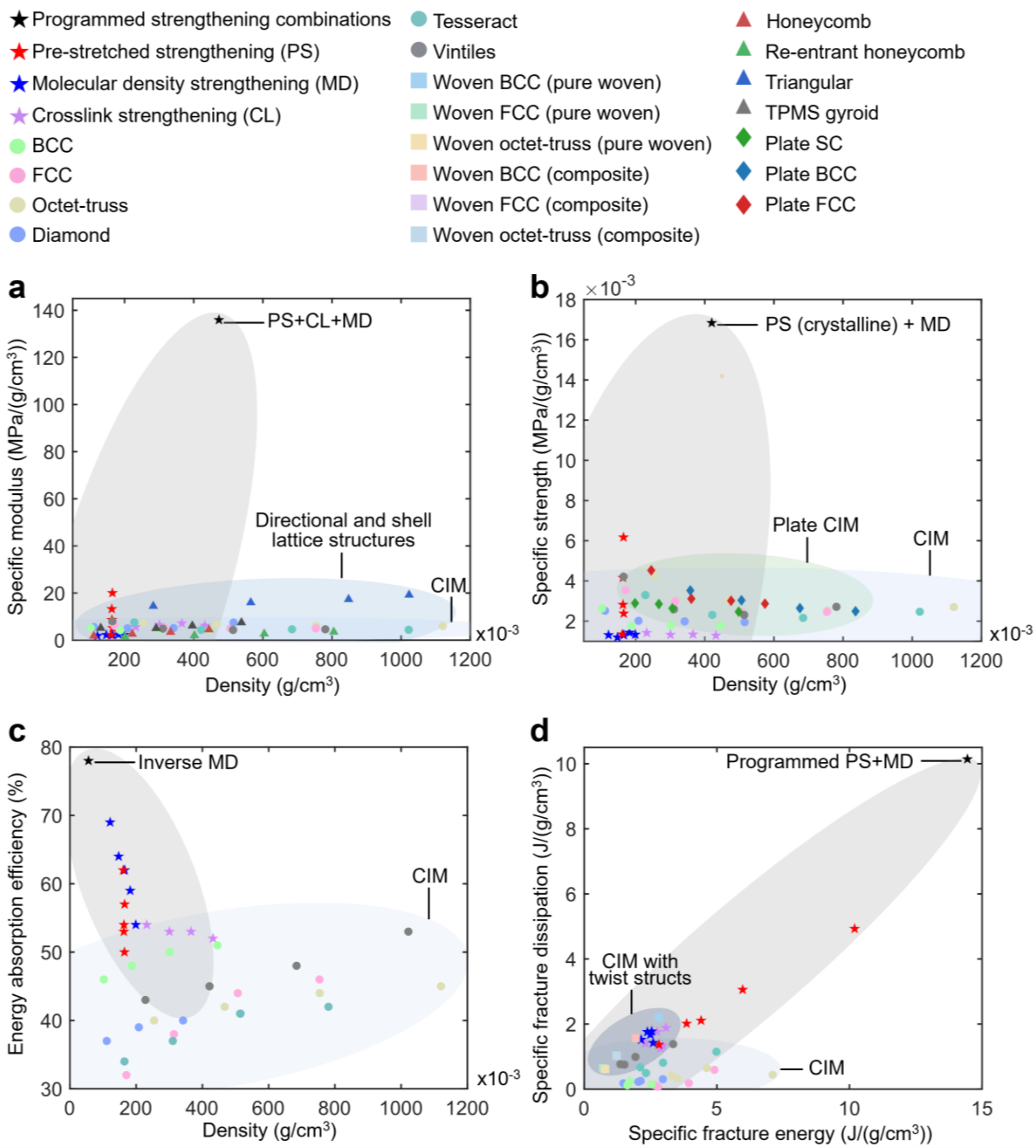

Fig. 3. Mechanical property comparison of the PIMs, existing CIMs, CIMs with twist structs, plate CIM, and directional lattice designs. Here, MD, PS, and CL denote MD strengthening, PS strengthening, and CL strengthening, respectively, while BCC, FCC, SC, and TPMS represent body-centered cubic, face-centered cubic, simple cubic, and triply periodic minimal surface. The parent material is Formlabs® durable resin. (a) The specific modulus versus density chart. (c) The specific strength versus density. (d) The compressive energy absorption efficiency versus density. (e) The specific fracture dissipation versus specific fracture energy.

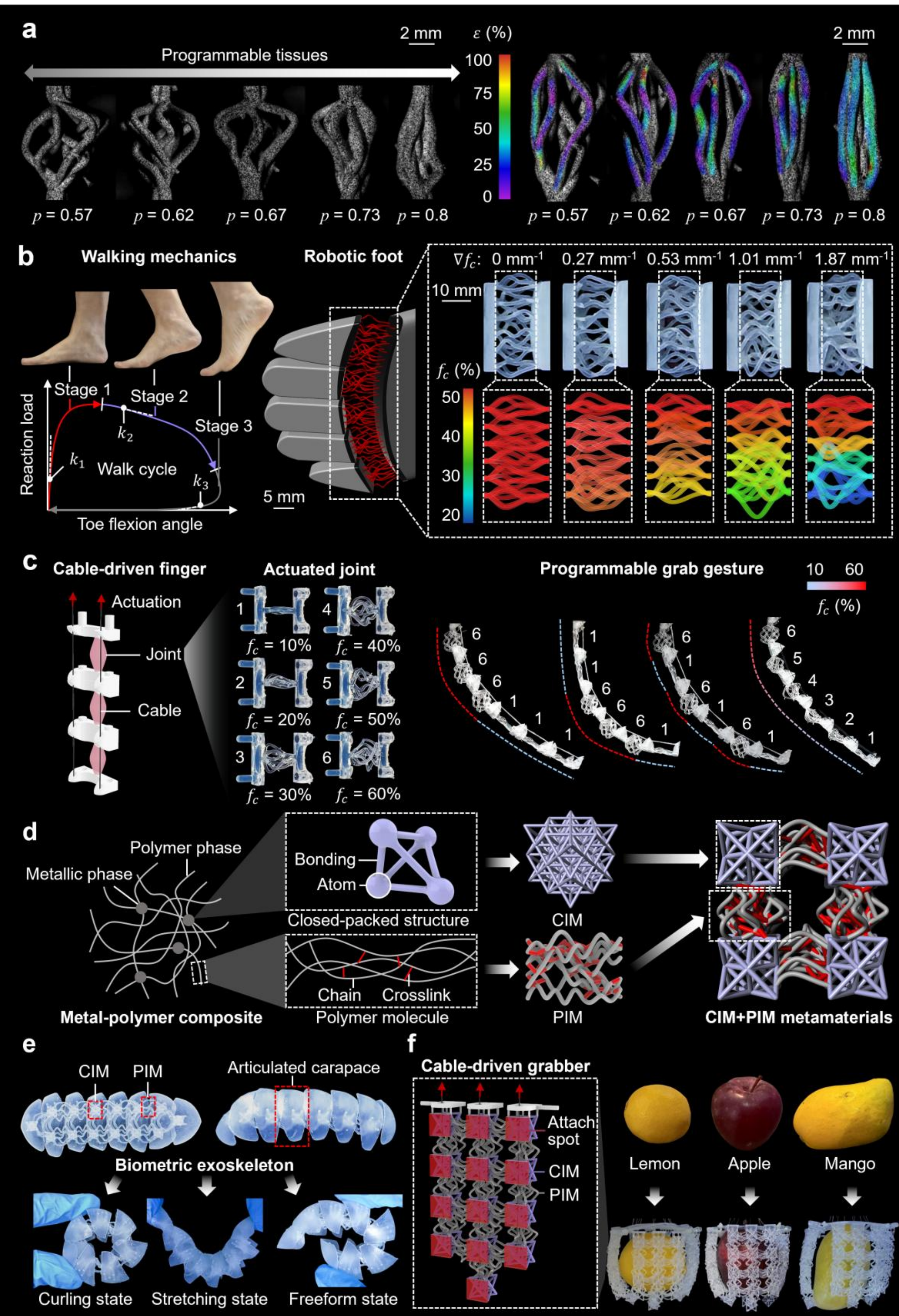

Fig. 4. The programmable PIM humanoid tissue and its functional applications. (a) The digital image correlation (DIC) analysis of the strain fields for PIM humanoid tissues with different pre-programmed PS levels. (b) Programmable tissue extensions per stress stimulus for the PIM humanoid tissues captured at 30% tissue strain. (c) Cable-driven finger enabled by PIM humanoid tissues at actuated joints. Here, 1 to 6 represents different actuated joints with $f_c$ engineered ranging from 10% to 60%, respectively. (d) Polymer + crystal lattice structure integration enabled by PIMs. (e) A biometric exoskeleton developed by CIM+PIM offering highly deformable shapes while preserving its protective functions. (f) A cable-driven grab finger with CIM+PIM that can grasp customized geometries.